\documentclass[useAMS,usenatbib]{mn2e}
\usepackage{times}
\usepackage{amsmath}
\usepackage{epsfig}

\title[Vigorous star formation in a bulge-dominated ERO at $z = 1.34$]{Vigorous star formation in a bulge-dominated ERO at $z = 1.34$}

\author[Garret Cotter, Chris Simpson \& Rosemary C. Bolton]
{
Garret Cotter $^{1,2}$\thanks{E-mail: garret@astro.ox.ac.uk}, Chris Simpson$^3$ and Rosemary C. Bolton$^{2}$\\
$^{1}$University of Oxford, Department of Astrophysics, Denys Wilkinson Building, Keble Road, Oxford OX1 3RH, UK\\ 
$^{2}$University of Cambridge, Department of Physics, Cavendish Laboratory, Madingley Road, Cambridge CB3 0HE, UK\\
$^{3}$Department of Physics, University of Durham, Rochester Building, Science Laboratories, South Road, Durham DH1 3LE, UK
}
\begin{document}

\date{Accepted version, 1 April 2005}

\maketitle

\begin{abstract}

We present near-IR spectroscopy of three Extremely Red Objects (EROs)
using the OHS/CISCO spectrograph at Subaru
telescope. One target exhibits a strong emission line which we
identify as H$\alpha$ at $z = 1.34$. Using new and existing ground-based
optical and near-IR imaging, and archival {\em HST} imaging, we argue
that this target is essentially an elliptical galaxy, with an old
stellar population of around $4 \times 10^{11} M_{\odot}$, but having
a dust-enshrouded star-forming component with a SFR of some
50--100$M_{\odot} {\rm yr}^{-1}$. There is no evidence that the galaxy
contains an AGN. Analysis of a further two targets, which do not
exhibit any features in our near-IR spectra, suggests that one is a
quiescent galaxy in the redshift range $1.2 < z < 1.6$, but that the
other cannot be conclusively categorised as either star-forming or
quiescent.

Even though our first target has many of the properties of an old
elliptical, the ongoing star formation means that it cannot have
formed {\em all} of its stellar population at high redshift.  While we
cannot infer any robust values for the star formation rate in
ellipticals at $z > 1$ from this one object, we argue that the
presence of an object with such a high SFR in such a small sample
suggests that a non-negligible fraction of the elliptical galaxy
population may have formed a component of their stellar population at
redshifts $z \sim$ 1--2. We suggest that this is evidence for ongoing
star formation in the history of elliptical galaxies.

\end{abstract}


\section{Introduction}

It has long been known that the properties of present-day ellipticals
and spiral bulges are well described by models in which the individual
bulge formed in a monolithic collapse at an early cosmic epoch, and
such models were favoured on grounds of simplicity (Eggen, Lynden-Bell
\& Sandage 1962).\nocite{1962ApJ...136..748E} However, in the
presently-favoured Cold Dark Matter (CDM) models of the Universe
(e.g., Spergel at al. 2003), the formation of galaxies is most
strongly influenced by the large galactic haloes of dark matter, and
theory now favours the hierarchical picture in which massive galaxies
are formed by the merger of smaller components. There has been
substantial success in reproducing observational results within this
framework (e.g., Fukugita, Hogan \& Peebles
1996)\nocite{1996Natur.381..489F}.

When the first Extremely Red Objects (EROs; e.g. Hu \& Ridgway 1994,
Graham \& Dey 1996, Smail et al. 2002)\nocite{1994AJ....107.1303H,
1996ApJ...471..720G, 2002ApJ...581..844S} were discovered,
considerable interest was aroused as it was thought that they were
passively-evolving galaxies which had formed monolithically at a very
early time. The situation was complicated when the first ERO to be
studied in detail, commonly known as HR10, proved to be an irregular
galaxy, slightly dust-reddened, undergoing an extreme starburst with a
$\sim 1000 M_{\odot} {\rm yr}^{-1}$ star formation rate (Cimatti et
al. 1998). The picture of the ERO population which has since developed
is that these red galaxies can be separated into two categories
(Mannucci et al. 2002; Pozzetti et
al. 2003)\nocite{2002MNRAS.329L..57M}\nocite{2003A&A...402..837P}.
The first consists of objects similar to HR10, which are shown by
high-resolution imaging to have disturbed morphologies. These galaxies
are clearly undergoing extreme star formation, sometimes contain an
AGN, and appear in many respects to be similar to the population of
submm-selected galaxies which are observed predominantly to be at
slightly higher redshifts, with a  median $z = 2.4$
(Chapman et al. 2003). \nocite{2003Natur.422..695C} A similar number
of EROs, however, appear to have the morphology and colours of
passively-evolving monolithically-formed ellipticals (although a very
small fraction of these, as evidenced by surveys for radiogalaxies,
also contain an AGN; see e.g. Willott at al. 2003). This presents two
problems: first, in resolving their star-formation history with CDM
hierarchical structure formation, and second, in the suggestion that
the numbers of faint red galaxies are inconsistent with models in which
the majority of today's ellipticals formed in a single burst of star
formation at very high redshift (Zepf
1996). \nocite{1997Natur.390..377Z}

To resolve this issue, it is necessary to measure accurately the rate
of star formation in the whole ERO population, not just the extreme
cases. This presents a major observational challenge, since the
H$\alpha$ emission line is redshifted into the 1--2 $\mu$m region of
bright night-sky airglow lines. An alternative diagnostic, the
[O{\sc~ii}] 3727-\AA\ line, is more heavily extinguished by the dust
which is commonly associated with star formation, making it less
reliable as a quantitative indicator. It too is affected by the rising
airglow forest and can only be reliably measured in a few small
redshift windows below $z=1.4$.  
Optical studies have, of necessity,
concentrated on the bluest (and hence optically brightest) EROs, which
could be biased in favour of those objects with recent star formation
(\nocite{2002A&A...392..395C}\nocite{2002A&A...391L...1C}
\nocite{2002A&A...384L...1D}e.g. Cimatti et al. 2002 a,b; Daddi et al. 2002). 
Studies of objects with
optical--infrared colours more typical of the general ERO population
are therefore necessary.  These arguments prompted us to use
the sensitive OH-suppression spectrograph at Subaru Telescope
\nocite{2002PASJ...54..315M} to obtain spectroscopic observations of
EROs in the crucial 1--2 micron region. In this paper we describe our
observations of a small sample of objects, concentrating on the
discovery that one of these, despite appearing a quiescent elliptical
in its morphology and near-IR colours, appears to have a considerable
degree of ongoing star formation.  Throughout this paper, we adopt
$H_0=70\rm\,km\,s^{-1}\,Mpc^{-1}$, $\Omega_{\rm m}=0.3$, and
$\Omega_\Lambda=0.7$, and all magnitudes are AB magnitudes.

\section{Observations and Analysis}

\subsection{Subaru OHS/CISCO spectroscopy}

Our initial target list was drawn from the ERO sample of Haynes (1998, hereafter H98;
see also Haynes et al. 2002 and Cotter et al. 2002), which was drawn from the $K$-band
imaging of Saunders et al. (1997) and $R$-band imaging from H98. The
sample was selected to have $R - K > 3$; for completeness we
include details of this sample in Table 1 and a diagram
of their distribution on the sky in Figure 1. From
this sample, we assembled a target list comprising the five objects
with $R - K > 4.0$.

\begin{figure*}
\label{toby_positions}
\epsfig{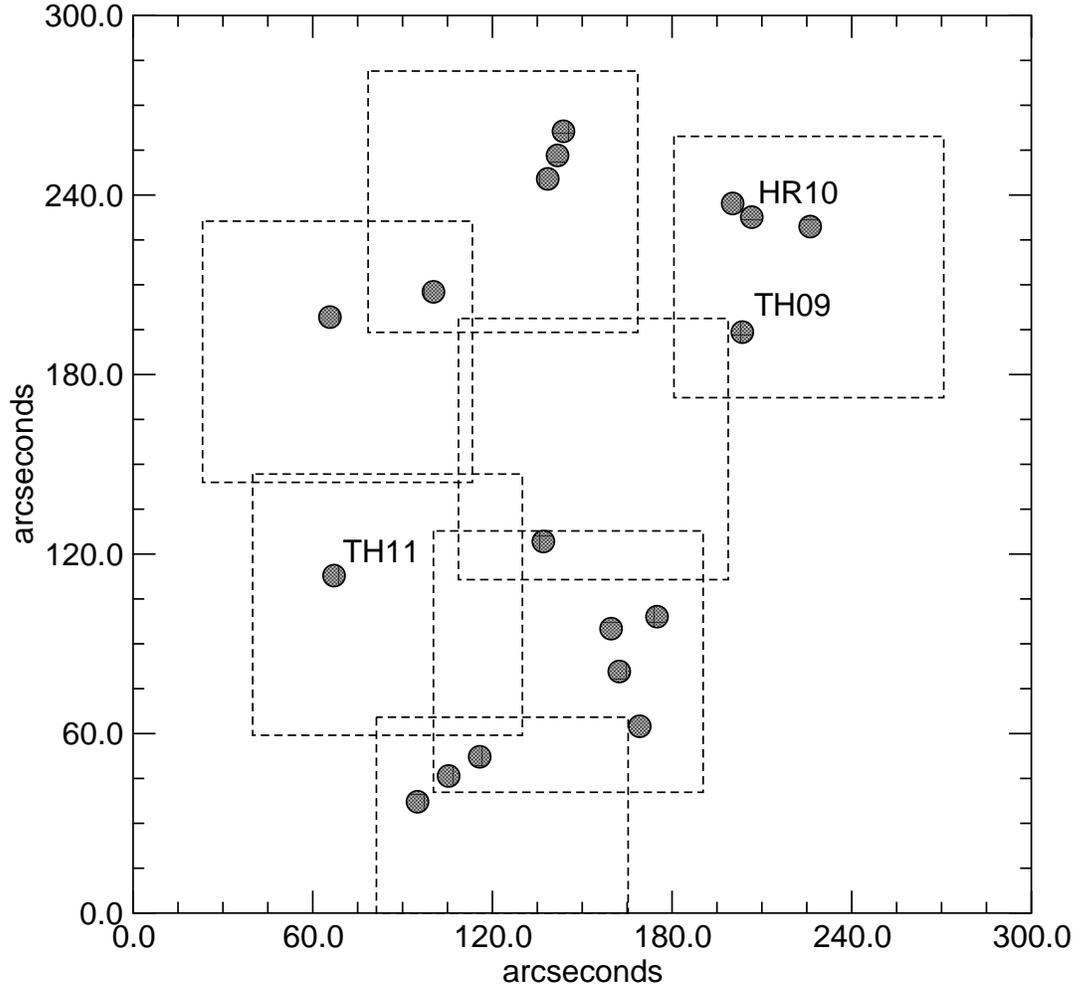}
\caption{Angular distribution of the 18 objects identified by Haynes (1998) to have $R - K > 3.0$ AB magnitudes. 
The dashed squares represent the area observed in $J$ and $K$ by Saunders et al. (1997).}
\end{figure*}

\begin{table*}
\label{toby_table}
\begin{tabular}{|l|l|l|l|l|}
\hline
ID   &  RA (J2000)  &      Dec  (J2000) &  	$R-K$ (AB)&	$K$(AB)  \\
\hline
\hline
TH01 & 16 45 10.24  & 	46 26 59.67  & 	4.811  	 &	20.62 	      \\
TH02 & 16 45 10.45  &	46 26 51.56  &	3.015  	 &	19.97          \\ 
TH03 & 16 45 10.77  &	46 26 43.72  &	3.176  	 &	20.55 \\
TH04 & 16 45 04.80  &	46 26 34.47  &	3.929  	 &	21.51  \\  
TH05 & 16 45 04.18  & 	46 26 29.74  &	3.048  	 &	22.02   \\ 
TH06 & 16 45 02.30  &	46 26 26.23  &	4.749  	 &	20.44$^{*}$ 	\\
TH07 & 16 45 14.53  &	46 25 56.24  &	3.105  	 &	20.72     \\
TH08 & 16 45 17.88  &	46 25 58.16  &	3.454  	 &	21.66$^{+}$ \\
TH09 & 16 45 04.55  &	46 25 51.34  &	4.088  	 &	19.84   \\
TH10 & 16 45 11.12  & 	46 24 42.46  &	3.338  	 &	21.15  \\
TH11 & 16 45 17.89  &	46 24 31.97  &	4.308  	 &	19.93  \\
TH12 & 16 45 07.49  & 	46 24 16.91  &	5.118  	 &	21.53    \\   
TH13 & 16 45 08.99  & 	46 24 13.11  &	3.578  	 &	22.15  \\
TH14 & 16 45 08.75  &	46 23 58.77  &	3.344  	 &	19.12  \\
TH15 & 16 45 08.12  &	46 23 40.35  &	3.347  	 &	20.77  \\
TH16 & 16 45 13.30  &	46 23 30.87  &	4.363  	 &	21.02  \\
TH17 & 16 45 14.31  &	46 23 24.59  &	3.044  	 &	20.76  \\
TH18 & 16 45 15.33  &	46 23 16.06  &	3.139  	 &	20.69  \\
\hline
\end{tabular}
\caption{The 18 objects identified by Haynes (1998) as having $R - K
> $~3.0 AB magnitudes. $^*$ TH06 is object HR10 from Hu \& Ridgway
(1994). $^+$ TH08 is a point source in the WFPC2 imaging and so we
classify it as a star.}
\end{table*}

Observations were made using the OH-airglow suppressor (OHS; Iwamuro et
al.\ 2001) and the Cooled Infrared Spectrograph and Camera for OHS (CISCO;
Motohara et al.\ 2002) at Subaru telescope on the night of UT 2001 June 11.
OHS suppresses the night-sky background in the $J$ and $H$ bands by
reflecting the dispersed beam from the target against a mirror that is
finely ruled to mask the wavelengths of individual OH lines.  The beam is
then recombined and dispersed once more by a lower-resolution grating, so
that the full spectrum may be projected onto the 1024$^2$-pixel HgCdTe
HAWAII detector of CISCO.

Unfortunately, although the night was photometric, observing
conditions were poor. High windspeed induced some telescope shake and
the seeing was between 1.0 and 1.5 arcseconds. This necessitated the
use of a 1.0-arcsecond slit, giving a spectral resolution
$R\approx200$. Nonetheless, we were able to obtain spectra of three
galaxies. Our first two targets are the brightest objects at $K$ from
our $R - K > 4.0$ list; for brevity we refer to these as TH09 and
TH11, as used by H98. We also obtained a spectrum of galaxy HR14 from
Hu \& Ridgway (1994), which fulfills our selection criterion of $R - K
> 4.0$ but is slightly outside the region observed by Saunders et al. (1997).
Coordinates and full photometry from the literature are presented in
Table 2 for all three targets.

For each target, individual 1000-s exposures (this length of exposure is
necessary to obtain a sky-limited spectrum in the OH-suppressed regions)
were offset by ten arcseconds in a standard ABBA pattern; TH09 and TH11
were observed for a total of 8000s each, and HR14 was observed for a total
of 4000s. Shorter observations were also made of the nearby stars SAO~46230
and SAO~46869 to permit correction of the atmospheric absorption. The data
were reduced using software written by CS which forms quads, flatfields the
images, removes residual sky emission, and combines the positive and
negative beams (see Simpson et al.\ 2004 for more details). Spectra were
extracted in a 1.0-arcsecond aperture, corrected for atmospheric
absorption, flux-calibrated from the acquisition images for each
target, and are shown in Figure 2.

\begin{figure*}
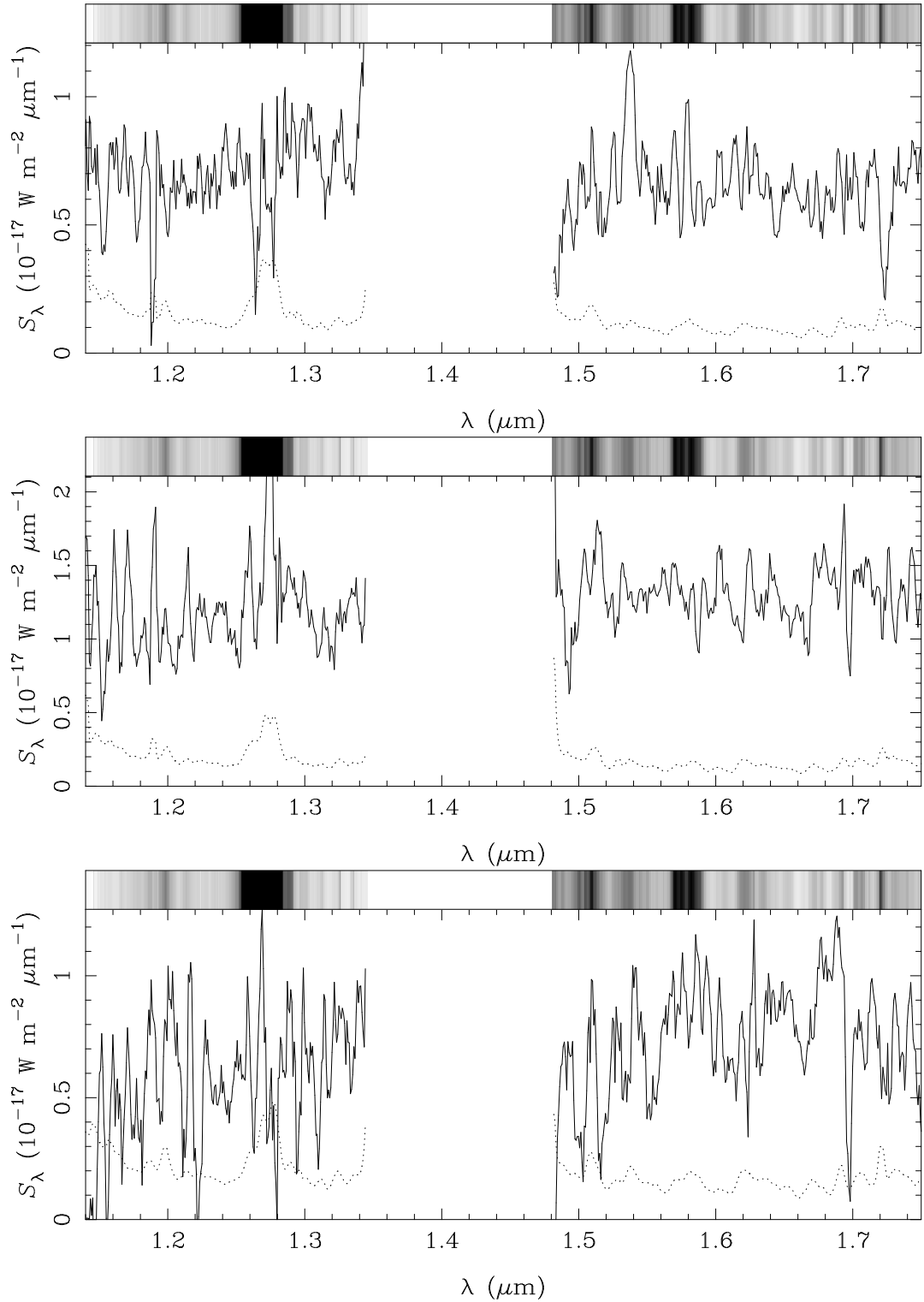

\label{spectra}
\psfig{figure=spek1.ps,angle=-90,width=0.8\linewidth}
\psfig{figure=spek2.ps,angle=-90,width=0.8\linewidth}
\psfig{figure=spek3.ps,angle=-90,width=0.8\linewidth}
\caption{Reduced spectra of the three OHS/CISCO targets (from top to
bottom: TH09, TH11, HR14). The spectra were extracted through a 1 arcsecond
aperture and smoothed using a five-pixel boxcar. The dotted line is the
estimated one-sigma error on the (smoothed) spectrum; the panel at the top
represents the relative intensity of the night sky after suppression of the
strongest OH emission lines.}
\end{figure*}
Only the spectrum of TH09 shows any significant feature: an emission line
at 15373 \AA\ is detected at high significance with a flux of $(4.9 \pm
0.8) \times 10^{-21}$ W m$^{-2}$ (${\rm EW} = 80$\,\AA), which is away from
any strong night-sky features. This line has a measured FWHM of
1350\,km\,s$^{-1}$ and is therefore unresolved spectrally. We identify this
line with H$\alpha$ 6563 \AA\ at $z = 1.34$. We reject the only other
plausible single-line identification, [O{\sc~ii}]~$\lambda$3727\,\AA\ at $z
= 3.12$, since we see no evidence for a 4000-\AA\ break or Balmer jump
despite the high signal-to-noise in the continuum, and this would imply an
absolute magnitude of $M_B=-25.0$ for the galaxy. We also consider an
alternative identification of [O{\sc~iii}]~$\lambda$5007\,\AA\ since our
upper limit to the presence of any $\lambda$4959\,\AA\ emission only rules
this out at the $2.5\sigma$ level. However, there is once again no evidence
for a 4000-\AA\ break or Balmer jump. More detailed SED fitting
(Section~3.1) is able to rule out these higher redshifts.

Having identified the line as H$\alpha$, we investigate the presence of
[N{\sc~ii}] emission. In the first instance, we remove the best-fitting
single Gaussian from the spectrum (since this line is unresolved, it should
not be a simple blend of H$\alpha$+[N{\sc~ii}]) and determine an upper
limit to the presence of an emission line at $\lambda_{\rm
rest}=6584$\,\AA. We also try to fit three emission lines simultaneously
using the {\tt specfit} package (Kriss 1994).  Both methods fail to provide
any evidence for [N{\sc~ii}] emission, with an upper limit of
[N{\sc~ii}]/H$\alpha < 0.25$ (95\,per cent confidence).

\subsection{UKIRT UFTI imaging}

$H$-band images of our three targets were taken with the UKIRT Fast Track
Imager (UFTI) on the UK Infrared Telescope on the nights of UT 2005
February 17 and 18. Conditions were photometric and the seeing was measured
to be about 1.0 arcseconds, comparable to that in which our spectroscopy
was taken. Nine-point jitter patterns with an exposure time of 60\,s per
pointing were used, with total exposure times of 3240\,s (TH09 and HR14,
which lie within a single UFTI field of view) and 2700\,s (TH11). The data
were reduced using the ORAC-DR pipeline (Economou et al.\ 1999) and
flux-calibrated from observations of FS~27 (Hawarden et al.\ 2001).
Photometry was performed in apertures of diameter 3 arcseconds and is
included in Table 2.

\begin{table*}

\label{our_targets}

\caption{Details of our spectroscopic targets, in the following
format: Coordinates; short name used in this paper; photometry (3-arcsec
apertures) in AB magnitudes from Haynes (1998), Hu \& Ridgway
(1994), or this work.}

\begin{tabular}{|l|l|l|l|l|l|l|l|l|l|l|l|}
\hline
\small

RA (J2000)     &  Dec (J2000)     & Name  & $U$            & $g$            & $V$            & $R$            & $I$           & $J$          & $H$          &  $K$          \\

\hline
16 45 04.5    & +46 25 51.34     & TH09  & $26.6\pm0.2$  & $25.0\pm0.64$  & $>24.8\pm0.1$  & $23.9\pm0.03$  & $22.7\pm0.03$ & $20.7\pm0.04$ & $21.0\pm0.1$ & $19.8\pm0.03$ \\

16 45 17.9    & +46 24 31.97     & TH11  & $26.8\pm0.2$     & $25.3\pm0.1$ & $25.532\pm0.2$ & $24.238\pm0.04$ & $22.9\pm0.05$ & $21.0\pm0.4$ & $20.9\pm0.1$ & $19.93\pm0.04$ \\

16 44 57.0    & +46 26 02.0      & HR14  & $29.0\pm0.3$ & $27.4\pm0.2$ & $27.7\pm0.2$ & $27.4\pm0.1$ & $25.6\pm0.1$  & $22.3\pm0.3$ & $21.5\pm0.1$ & $20.6\pm0.2$ \\
\hline

\end{tabular}

\end{table*}

\subsection{{\em HST} WFPC2 imaging}

Archival Wide-Field/Planetary Camera 2 (WFPC2) observations are available
for TH09 and HR14. TH09 was observed for a total of 3000s in F555W and
6000s in F814W, using a four-point sub-pixel dither pattern, as part of PID
7342 (PI: R. Saunders). These were processed using the {\tt drizzle}
package (Fruchter \& Hook 2002) \nocite{2002PASP..114..144F} to recover
some of the resolution of the undersampled WFPC2 PSF. HR14 was observed for
a total of 5300s in F814W as part of PID 6598 (PI: A. Dey) and these data
were processed using standard {\sc iraf} procedures. Greyscale close-ups of
the targets are presented in Figure 3.

\begin{figure*}
\label{wfpc_images}

\begin{tabular}{cc}
\parbox{9cm}{
\psfig{figure=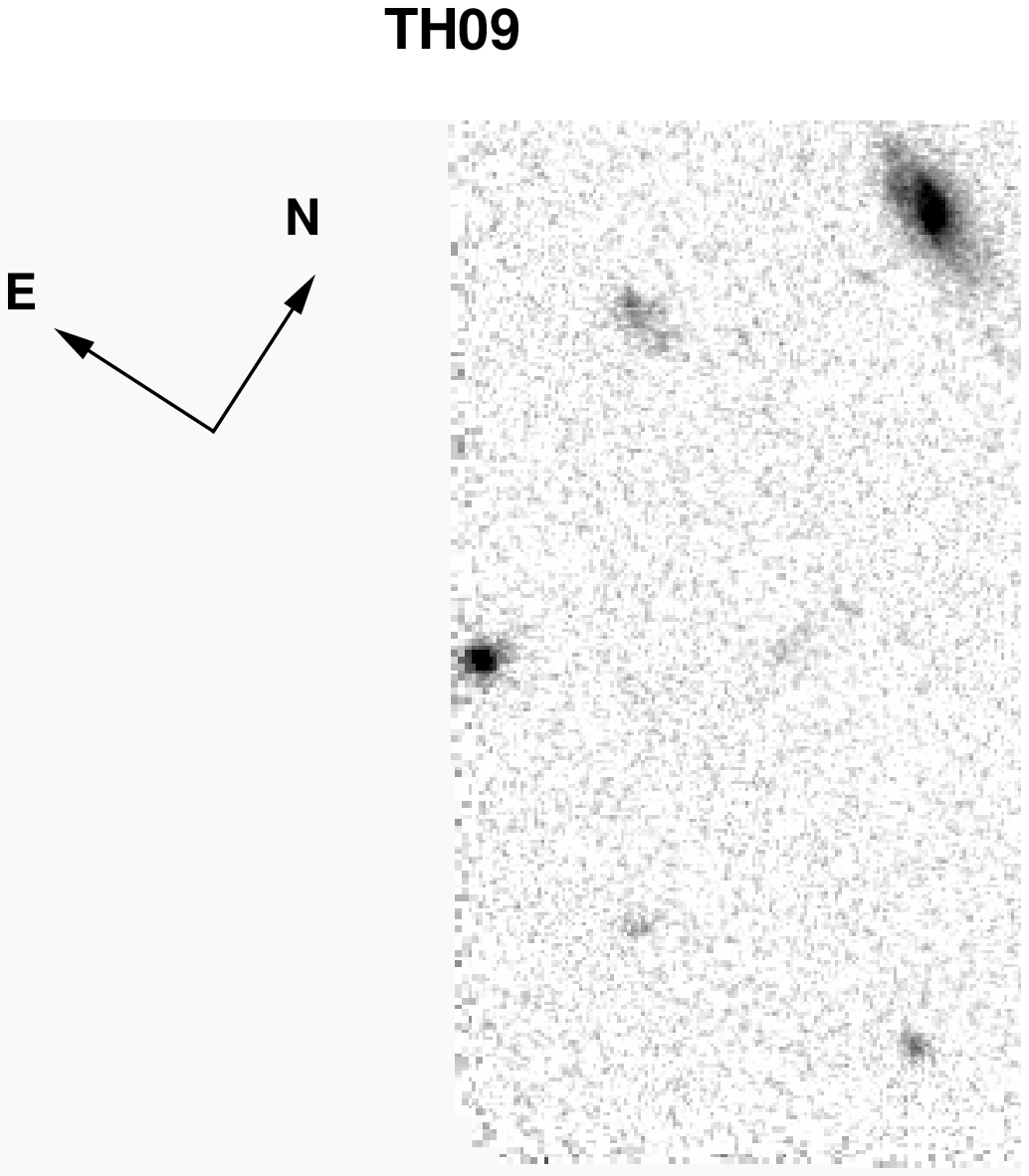,width=0.9\linewidth,angle=0}
}
\parbox{9cm}{
\psfig{figure=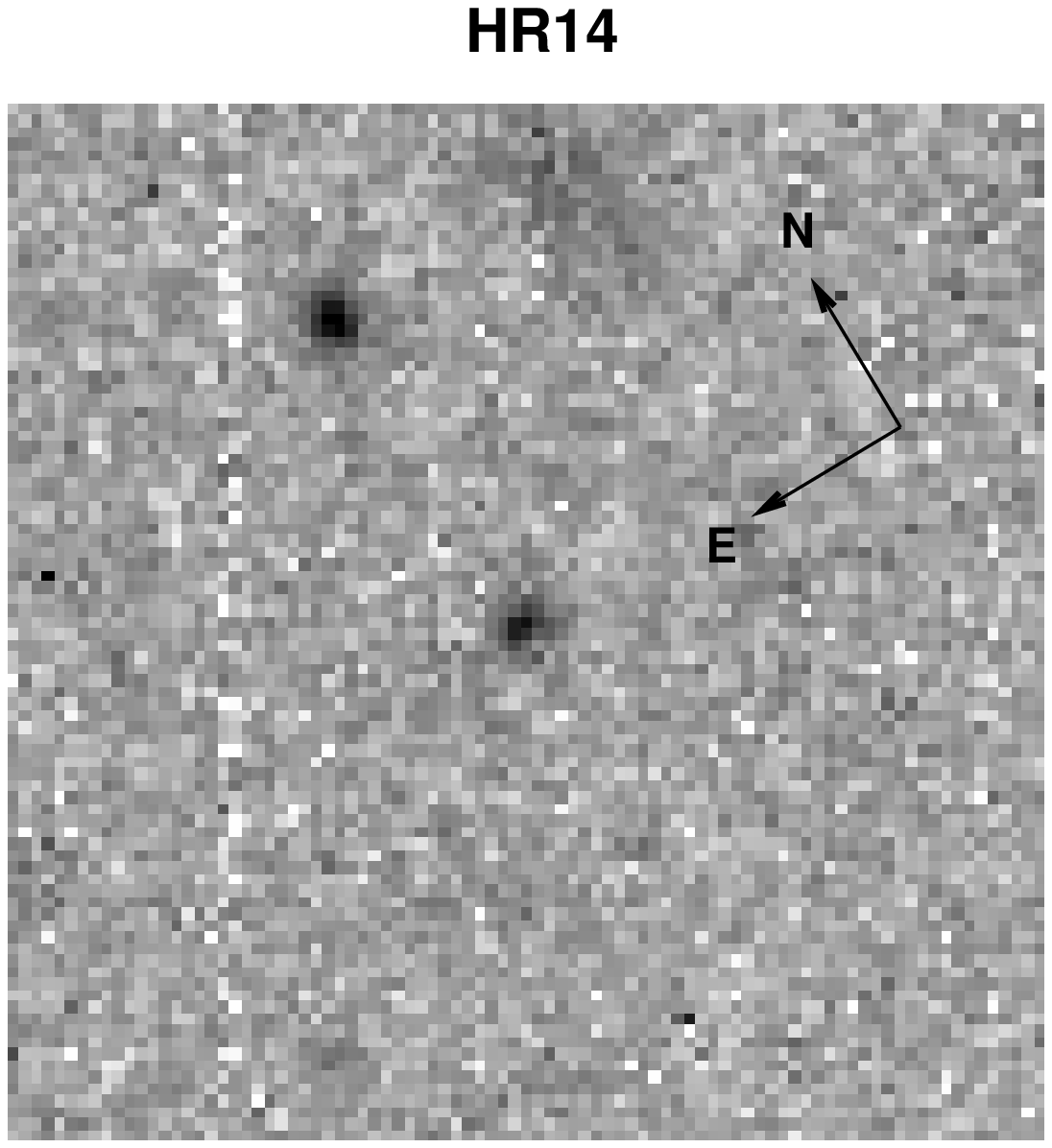,width=0.9\linewidth,angle=0}
}
\end{tabular}
\caption{{\em HST} WFPC2 F814W images of TH09 (left) and HR14
(right). The greyscale is logarithmic and each image is 18 arcseconds
square.  Note that TH09 is close to the edge of the WFPC chip, and so
the un-drizzled pixels appear at the usual WFPC pixel scale. Data from
the edge of the field was not used in the GIM2D fit. The image of HR14 is not drizzled. }
\end{figure*}

The GIM2D package (Simard 1998) was used to fit the two-dimensional
surface brightness profiles of TH09 and HR14 in the WFPC2 images. We
considered bulge-only fits, disc-only fits, and fits containing both
bulge and disc components. Only the fit to TH09 displayed a clear
morphological preference, with the bulge-to-total luminosity ratio of
the best fit being $b=0.75.\pm 0.13$ (95 per cent
confidence), and there being no significant preference for this fit
over the pure bulge fit (reduced $\chi^2/\nu$ of 1.18 and 1.20,
respectively, compared to 1.33 for a pure disc fit).

\begin{figure*}
\label{gim2d}
\psfig{figure=TH09_gim2d.ps,width=0.3\linewidth,angle=-90,clip=}
\psfig{figure=HR14_gim2d.ps,width=0.3\linewidth,angle=-90,clip=}
\vskip 0.5 cm
\psfig{figure=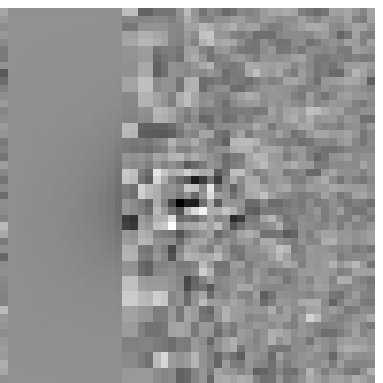,width=0.3\linewidth,clip=}
\hskip 2cm
\psfig{figure=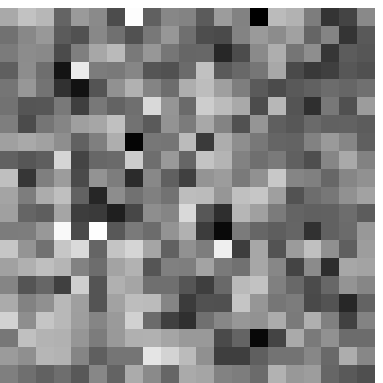,width=0.3\linewidth,clip=}
\caption{Top: {\sc GIM2D} fits for TH09 (left) and HR14
(right). Points with error bars are the data; solid line is best
bulge-plus-disc fit, dashed bulge only, dotted is disk only. Bottom:
greyscale close-ups of TH09 (left) and HR14 (right) with the best-fit
{\sc GIM2D} models subtracted.}
\end{figure*}

The shorter wavelength
of the F555W image of TH09 might be expected to increase the contribution
of any disc component, but the lower signal-to-noise ratio of this image
precludes a preference being given to any particular model because the fits
are poorly constrained (giving a reduced $\chi^2/\nu \approx 0.4$). We note
however that the preferred composite model has a strong bulge component (93
per cent of the total luminosity). The best fit for HR14 has a bulge
contribution of 32 per cent, but the poor sampling and relatively low
signal-to-noise ratio mean that all three fits have very similar values of $\chi^2$. 
We show the results of our fits in
Table~\ref{gim2dtab} and Figure~\ref{gim2d}.

\begin{table*}
\begin{tabular}{llccccccc}
Galaxy & Fit & $m_{\rm tot}$ & B/T & $\mu_{\rm b}$ & $r_{\rm b}$ &
$\mu_{\rm d}$ & $r_{\rm d}$ & $\chi^2/\nu$ \\ 
& & (mag) & & (mag arcsec$^{-2}$) & (arcsec) & (mag arcsec$^{-2}$) &
(arcsec) \\
\hline
& Bulge-only & $22.42\pm0.05$ & 1.0 & $15.65\pm0.19$ &
$0.43^{+0.05}_{-0.04}$ & -- & -- & 1.20 \\
TH09 & Disc-only & $22.89\pm0.04$ & 0.0 & -- & -- & $20.45\pm0.17$ &
$0.13\pm0.01$ & 1.33 \\
& Bulge+disc & $22.59\pm0.07$ & $0.75\pm0.13$ & $15.80\pm0.35$ &
$0.37^{+0.08}_{-0.05}$ & $21.83\pm0.43$ & $0.14\pm0.03$ & 1.18 \\
\hline
& Bulge-only & $24.35\pm0.36$ & 1.0 & $16.79^{+1.17}_{-1.41}$ &
$0.30^{+0.21}_{-0.15}$ & - & -- & 0.88 \\
HR14 & Disc-only & $24.71\pm0.22$ & 0.0 & -- & -- &
$21.71^{+0.51}_{-0.58}$ & $0.10\pm0.02$ & 0.88 \\
& Bulge+disc & $24.56\pm0.31$ & $0.32^{+0.39}_{-0.32}$ &
$19.07\pm1.83$ & $0.44^{+0.37}_{-0.44}$ & $21.98\pm0.95$
& $0.10\pm0.03$ & 0.88 \\
\hline
\end{tabular}
\caption{Best-fit parameters for the fits shown in
Fig.~\ref{gim2d}. The quoted errors are 99\% confidence intervals. The
confidence intervals for the HR14 bulge+disk fit are
approximate.\label{gim2dtab}}
\end{table*}

We thus conclude that TH09 is morphologically a bulge-dominated
galaxy, and is likely a pure elliptical, but that we are unable on the
basis of these images to make a robust morphological classification
for HR14.

\section{Discussion}

\subsection{Detailed properties of TH09}

We first concentrate our discussion on TH09. As noted in the previous
section, this galaxy is morphologically an elliptical. To investigate
the properties of the stellar population in TH09 we first used the
{\sc Hyperz} photometric code (Bolzonello, Mirales \& Pello, 2000),
fixing the redshift at 1.34. This gave a poor fit for any of the model
spectra in {\sc Hyperz}; this also applied to the potential {\sc
[Oii]} and {\sc [Oiii]} redshifts described in section 2.1. However,
on examining the best-fit spectrum for $z = 1.34$ in detail, it became
apparent that the observed-frame near-IR was in fact a very good fit
to an old population. The poorness of the overall fit was caused by
the rest-frame UV being too bright for this population, by about a
factor of ten at $g$-band.

We next considered the possibility that TH09 hosts an AGN.
The galaxy is not detected to a three-sigma limit of 18 $\mu$Jy at 8
GHz (Cotter, unpublished data) and a deep $ROSAT$ PSPC image (Kneissl,
Sunyaev \& White 1997) gives a limit of $4 \times 10^{-14}$ erg
s$^{-1}$ cm$^{-2}$ in the 0.5-2.0 keV band. Although there are no deep
Chandra or XMM images available, which would allow us to rule out hard
X-ray emission, these non-detections argue strongly against TH09 being
an AGN host.

Our detection detection of H$\alpha$ therefore implies that TH09 is
starforming; this means that TH09 cannot be quiescent old elliptical
galaxy with {\em all} its stars having formed at an early cosmic
epoch. We chose then to model the SED of TH09 as a combination of an
old stellar population plus a young starburst population. We used the
solar metallicity Padua 1994 SSP from Bruzual \& Charlot (2003;
hereafter BC03), with a model age of 5.5 Gyr (the age of our fiducial
universe at $z = 1.34$) for the old population, as this gave the best
fits to the red end of the SED (lower metallicity SSPs were not
sufficiently red at this age). For the young stellar component we used
a continuous starforming model from the same SSP with an age of 70 Myr
(at which are the shape of the SED becomes close to its asymptote),
using the rest-frame 2800 \AA flux to obtain the SFR.

The next step was to estimate the SFR from the H$\alpha$ luminosity in
our OHS/CISCO spectrum.  Since our spectral extraction aperture
encompasses only a fraction of the total flux from the galaxy, we
scale our H$\alpha$ flux to account for slit losses. We measure a
difference of one magnitude between the brightness of TH09 in our
extraction aperture (as determined from the acquisition images) and
its flux in Table 2, and therefore apply a correction
factor of 2.5 to our measured emission line flux. This implicitly
assumes that the H$\alpha$ line emission has the same angular extent
as the $H$-band light, although we can determine a robust lower limit
to the size of this correction of $>1.5$ by determining how much light
from a point source would fall into our spectroscopic aperture, and
hence our correction will not be in error by more than 40\,per
cent. Using the relationship between H$\alpha$ luminosity and star
formation rate given by Kennicutt (1998), this implies a star
formation rate of $\sim 20 M_{\odot}\,{\rm yr}^{-1}$. However, adding
a suitable constant star-forming component to our model spectrum
produces too much emission in the rest-frame ultraviolet. At this
point we note that, although there is a clear rest-frame UV excess in
the spectrum of TH09, it does turn over in the far UV, so we also
included a dust extinction, using the reddening law of Calzetti et
al. (2000) which we applied only to the young stellar population.

We find that by taking an unreddened old stellar population, and
adding a young population commensurate to a star formation rate of
50-100 $M_{\odot} {\rm yr}^{-1}$, with dust extinctions of $A_V =
3$--4, we obtain reduced $\chi^2$ values of around 3 (in
this region of the parameter space, the slope of the $\chi^2$ surface
is quite shallow around the minimum). Moreover, this rate of star
formation is then consistent with  the observed H$\alpha$ flux,
assuming that it originates only in the star forming regions and
suffers the same extinction.

The limit to the 8.4-GHz continuum flux density implies a
three-sigma upper limit on the SFR of 150 $M_{\odot} {\rm
yr}^{-1}$. To calculate this SFR limit, we use the prescription of
Haarsma et al. (2000),
\begin{equation*}
SFR  = Q\times{\frac{L_\nu}{W Hz^{-1}} /[5.3\times 10^{21}(\frac{\nu}{GHz})^{-0.8} \\ 
 +  5.5\times 10^{20}(\frac{\nu}{GHz})^{-0.1}}]
\end{equation*}
Here ${Q}^{-1}$ gives the fraction of mass in forming stars that goes
into stars of mass $M_{\odot}\ge 5$ $M_{\odot}$. A value of 5.5
was used for $Q$, appropriate for a Salpeter IMF with a mass range of 0.1
$M_{\odot}$ to 100 $M_{\odot}$. Note that the high-frequency flux is
dominated by the flat-spectrum thermal bremsstrahlung component, so
the fact that our limit is at a rest-frame frequency of 19 GHz should
not cause a large error in the inferred SFR.

We were then able to calculate the mass of old stars in TH09 via the
old stellar component in our best-fit model, using the luminosity from
1$M_{\odot}$ provided in the BC03 models. This gives a stellar mass
for the old population of approximately $4 \times 10^{11} M_{\odot}$.
TH09 is clearly a massive galaxy, at the upper end of the masses
observed in the FDF/GOODS-S survey (Drory et al. 2005) and also comparable to
the hosts of radiogalaxies in this redshift range (Willott et
al. 2003).

\begin{figure*}
\label{model_spec}
\epsfig{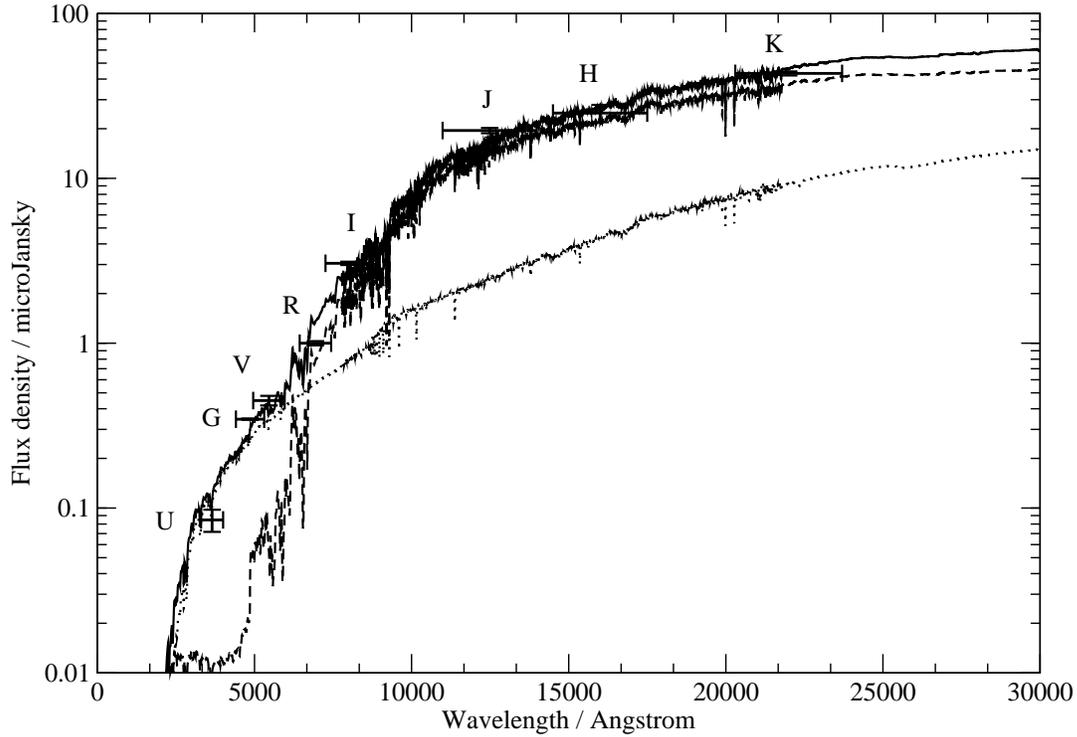}
\caption{The best-fit combined-population model for TH09, with a star
formation rate of 100 $M_{\odot}\,{\rm yr}^{-1}$, is shown by the
solid black line, along with the young (dotted) and old (dashed)
contributions, and the flux densities from the broad-band images. Note
that the young population makes a negligible contribution to the
observed-frame near-IR colours; the young population is demanded by
the excess flux at $g$-band}

\end{figure*}

\begin{figure*}
\label{rest_uv}
\epsfig{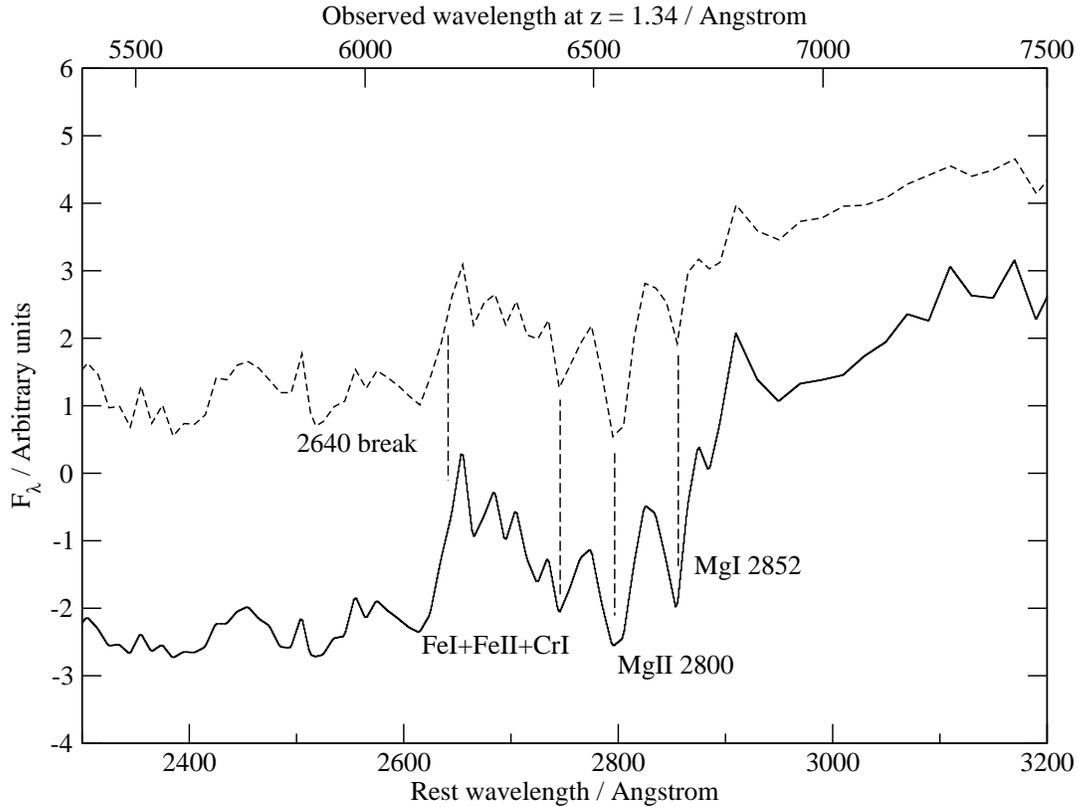}
\caption{Close-up of the rest-frame UV region of our model spectrum
for TH09 (solid line), along with a 1.1-Gyr old BC03 model (dashed
line), as used by Cimatti et al. (2004), shifted downwards by four
units for clarity. Note that in the spectral features here are
indicative of a relatively old population; spectroscopy in this
region---the observed-frame optical---cannot conclusively exclude
ongoing star formation.}
\end{figure*}

Despite the substantial mass of stars formed at an earlier epoch, we
predict that the star formation activity in TH09 would produce an
850-micron flux of $\sim1$\,mJy (using, e.g., the conversion between submm
flux and star formation rate of Ivison et al.\ 2002). \nocite{2002MNRAS.337....1I} This makes it
comparable to, and indeed perhaps representative of, the objects which
contribute to the faint end of the submm number counts (e.g. Hughes et al.\
1998; Barger et al. 1998; Smail et al.\ 1999\nocite{1998Natur.394..241H},
\nocite{1998Natur.394..248B} \nocite{1999MNRAS.308.1061S}).

We also compare our observations with the ``pixel-by-pixel''
photometry of galaxies in the Hubble Deep Field, in which many
ellipticals at redshifts slightly below $z = 1$ show evidence of
stellar populations of slightly different ages seemingly threaded
through the spheroids (Menanteau et al. 2001; Abraham et
al. 1999\nocite{2001MNRAS.322....1M}\nocite{1999MNRAS.303..641A}). We
suggest that TH09 is an elliptical galaxy in the process of an event
which will lead to such an appearance at a later epoch.

We note that in our model spectrum the region around Mg{\sc~ii}
2800 \AA\ is similar to the ERO spectra observed by Cimatti et
al. (2004) with FORS at the VLT (see Fig. 5).  The spectroscopic
sample of Cimatti et al. clearly demonstrates that there is a
substantial population of EROs which, at $1 < z < 2$, have already
built up a large mass of old stars. However, in a scenario where the
star-forming regions are obscured but the old population is not, the
spectral features in the observed-frame optical do not yield immediate
evidence of star formation. The implication here is that while elliptical
EROs do have a well-established old stellar population, optical
spectroscopy cannot rule out the possibility of ongoing star formation
in these galaxies.

Finally, in terms of its environment, Fig. 1 shows that TH09 appears to be a
member of a small group of four objects on the sky. However, since one
of these is HR10 at a significantly higher redshift, we caution
against concluding that TH09 is definitely in a rich environment.

\subsection{TH11 and HR14}

With no clear spectral features seen in TH11 and HR14, we used {\sc
Hyperz} to estimate the utility of our data for {\em ruling out\/}
star formation via H$\alpha$ non-detection.  First, we checked to
determine if, despite the dual-population of TH09, {\sc Hyperz} could
recover its spectroscopic redshift. We found most likely redshift $z =
1.30$, from a single-burst old population but with a poor fit. From
inspection it of the best-fit model it appears that the {\sc Hyperz}
redshift is approximately correct because of the colours in the
near-IR, especially around the rest-frame 4000-\AA\ break.

For TH11, the best fit model was again an old, unreddened single-burst
population, with a most likely redshift of $z=1.4$ and a 99 per cent
confidence range $1.2 < z < 1.6$. Our spectrum would be sensitive to
H$\alpha$ emission in the redshift range $1.28 < z < 1.66$, which
encompasses about 85 per cent of the {\sc Hyperz} redshift
probability. The lack of observable H$\alpha$ emission in our spectrum
(the $3\sigma$ upper limit is $\sim 2.3 \times 10^{-21}$\,W\,m$^{-2}$
over most of our spectrum) is therefore most likely to be due to a
real absence of emission (rather than the line falling between the $J$
and $H$ bands) and implies an unobscured star formation rate of $<15
M_\odot\rm\,yr^{-1}$. We also that this object falls marginally in the
``ellipticals'' region of colour-colour space defined by Pozzetti \&
Mannucci (2000).

For HR14, the situation is somewhat more complex, with a bimodal
likelihood distribution in redshift. HR14 is strikingly red, even for
an ERO; from Haynes et al. (2002) and Hu \& Ridgway (1994) we
calculate $R - K = $ 6.7 AB magnitudes.  An unreddened single burst
model provides a most likely redshift of $z = 2.5$, with a 99 per cent
confidence range of $2.1 < z < 3.2$; this is in agreement with Hu \&
Ridgway's estimate based on their $IJHK$ photometry.  However, nearly
equally-likely fits are obtained with a redenned populations at a
lower redshift: an old single-burst model with $A_V = 2.5$ has most
likely $z = 1.8$, 99 per cent confidence range $1.2 < z < 2.2$; and a
young population with $A_V = 4.0$ has most likely $z = 1.6$, 99
percent confidence range $1.27 < z < 1.89$.

It is certainly more physically plausible that if HR14 were such a
dusty object, it would likely also be strongly starforming. We would
then expect to see H$\alpha$ emission in our spectrum over most of the
likely redshift range, but the signal-to-noise in our spectrum is
about a factor of two worse than our other spectra. Considering also
our poor constraint on the bulge-to-disc fit for HR14, we cannot
conclusively decide if this is a starforming object or a more
quiescent galaxy at a high redshift (we note that Pozzetti \& Mannucci
classify HR14 as an ``elliptical'', although within the errors of the
photometry it does not lie unambiguously in this region of their
colour--colour diagram).

\subsection{Implications for the rest of the ERO population}

Our central observational conclusion is that in this small sample, one
of our targets is unambigously starforming, despite its morphological
appearance as a well-formed elliptical and the clear indication from
its near-IR colours that the bulk of its stellar population is not
dust-enshrouded. With our data we cannot determine whether TH09 is a
undergoing a very short-lived episode of star formation, or whether is
it undergoing a process wherein the old stellar population is being
augmented by a longer episode of star formation, perhaps by a
substantial increase in its cold gas reservoir via a merger.

However, we stress that the presence of an object such as TH09 in such
a small observation sample is indicative that such objects must
represent a non-negligible fraction of the ERO population. If this is
indeed that case, it would provide support for a picture in which
ellipticals frequently do not form in a single monolithic
star-formation. 

To determine precisely the rate of this continuing growth in the
stellar mass of the elliptical galaxy population, it will be essential
to determine spectroscopically which of the EROs are truly quiescent,
which host an AGN, and which are undergoing star formation. Such
observations are ideally suited to Subaru Telescope's second
generation instrument FMOS (Maihara et al.\ 2000; Kimura et al.\
2003), a wide-field, fibre-fed, OH-suppression spectrograph which will
be able to simultaneously take up to 400 spectra across a 30-arcminute
field over the crucial 0.9--1.8 \micron\ range.

\section{Summary}

We have presented new OHS/CISCO near-IR spectroscopy and an analysis
of new and existing HST and ground-based imaging of a small sample of
EROs. Our conclusions are as follows:

\begin{enumerate}

\item We have detected H$\alpha$ emission from one target, TH09, at
$z = 1.34$.

\item From an analysis of the broad-band colours of TH09, we
conclude that it contains an old, unreddened stellar population and
also a dusty young starforming component, with a star-formation rate
of some 50--100$M_{\odot} {\rm yr}^{-1}$. 

\item Analysis of archival {\em HST} imaging of this object
demonstrate that it is bulge-dominated, essentially an elliptical galaxy.

\item We detect no emission lines in our spectroscopy of two other
targets, TH11 and HR14. Using a photometric redshift analysis of the
broad-band colours of these two objects, we conclude that TH11 is
unlikely to be undergoing any significant star formation, but we are
unable to make any robust conclusions about the nature of HR14.

\item Given the appearance of an object such as TH09 in such a small
sample, we conclude that a non-negligible fraction of elliptical EROs
must be undergoing continuing star formation. While it remains clear
that these objects have a well-established old stellar population, we
suggest that models in which ellipticals form in their entirety in a
monolithic burst of star formation are oversimplified. We note that
future sensitive wide-field near-IR spectrographs will allow
measurement of the star formation rate in large numbers of such
objects.

\end{enumerate}

\nocite{2003SPIE.4841..974K}
\nocite{2000SPIE.4008.1111M}
\nocite{2000MNRAS.317L..17P}
\nocite{2004Natur.430..184C}
\nocite{2000ApJ...544..641H}
\nocite{1998ApJ...498..541K}
\nocite{2000ApJ...533..682C}
\nocite{2003MNRAS.344.1000B}
\nocite{1998PhDT........93H}
\nocite{1999ASPC..172...11E}
\nocite{2001MNRAS.325..563H}
\nocite{1994ASPC...61..437K}
\nocite{1997ApJ...479L...5S}
\nocite{2005ApJ...619L.131D}
\nocite{2002A&A...392..395C}
\nocite{2002MNRAS.334..262H}
\nocite{2002MNRAS.334..283C}
\nocite{2003ApJS..148..175S}
\nocite{1998Natur.392..895C}
\nocite{2003MNRAS.339..173W}
\nocite{1998ASPC..145..108S}
\nocite{2000A&A...363..476B}
\nocite{1998MNRAS.297L..29K}

\section*{Acknowledgments}

We warmly thank the referee for helpful comments on our first draft,
and Katherine Inskip and Steve Rawlings for useful discussions.

This work is based in part on data collected at Subaru Telescope,
which is operated by the National Astronomical Observatory of Japan;
and in part on observations made with the NASA/ESA Hubble Space
Telescope, obtained from the data archive at the Space Telescope
Science institute. STScI is operated by the Association of
Universities for research in Astronomy, Inc., under NASA contract NAS
5-26555. GC acknowledges support from PPARC grants KKZA/014 RG35306
and KKZA/01 RG32761. CS acknowledges a PPARC Advanced Fellowship. RCB
acknowledges a PPARC Ph.D. research studentship.

It is a pleasure to express our gratitude and respect to the
indigenous people of Hawai`i, on whose sacred mountain Mauna Kea our
infrared observations were made.

\def\apjs{ApJS}
\def\mnras{MNRAS}
\def\nat{Nature}
\def\aap{A \& A}
\def\apj{ApJ}
\def\apjl{ApJL}
\def\pasp{PASP}
\def\aj{AJ}
\def\pasj{PASJ}


\end{document}